\documentstyle[12pt]{article}
\begin{document}

\setcounter{footnote}{0}
 

\newcommand{\be}{\begin{equation}}
\newcommand{\ee}{\end{equation}}
\newcommand{\ba}{\begin{eqnarray}}
\newcommand{\ea}{\end{eqnarray}}
\newcommand{\bas}{\begin{eqnarray*}}
\newcommand{\eas}{\end{eqnarray*}}

\newcommand{\nn}{\nonumber}

\newcommand{\g}{\gamma}
\newcommand{\Lam}{\Lambda}
\newcommand{\lam}{\lambda}
\newcommand{\th}{\theta}
\newcommand{\dl}{\delta}

\newcommand{\bomega}{\mbox{\boldmath $\omega$}}

\newcommand{\tr}{\mbox{tr}}

\newcommand{\oh}{\frac{1}{2}}

\newcommand{\cA}{{\cal A}}
\newcommand{\cB}{{\cal B}}
\newcommand{\cC}{{\cal C}}
\newcommand{\cD}{{\cal D}}

\newcommand{\cP}{{\cal P}}
\newcommand{\cS}{{\cal S}}

\addtolength{\baselineskip}{0.20\baselineskip}
\hfill UT-Komaba 98-11
 
\hfill June 1998
\begin{center}
 
\vspace{36pt}
{\large \bf Integrable Impurity Model with Spin and Flavour: \\ 
Model Inspired by Resonant Tunneling in Quantum Dot}
 
\end{center}

\vspace{36pt}
 
\begin{center}

\vspace{5pt}
 Osamu Tsuchiya \footnote{E-mail address: otutiya@hep1.c.u-tokyo.ac.jp} \\
\vspace{3pt}
 Ikuo Ichinose \footnote{E-mail address: ikuo@hep1.c.u-tokyko.ac.jp} \\
\vspace{3pt}
Yasuyuki Kayama \footnote{E-mail address: kayama@amon.c.u-tokyo.ac.jp} \\
\vspace{3pt}
{\sl Institute of Physics, 
University of Tokyo, \\
Komaba, Meguro-ku, Tokyo 153-8902, Japan}

\end{center}

 
\vspace{36pt}
 
 
\begin{center}
{\bf Abstract}
\end{center}
 
\vspace{12pt}
 
\noindent
We introduce an integrable impurity model in which both electrons
and impurity have spin and flavour degrees of freedom.
This model is a generalization of the multi-channel Kondo model
and closely related with resonant tunneling through quantum dot.
The Hamiltonian is exactly diagonalized by means of the Bethe ansatz.

\vspace{24pt}
 
\vfill
 
\newpage
 
\setcounter{footnote}{0}

\section{Introduction}
Kondo type integrable impurity models have been studied extensively
since the discovery of  the exact solution to the single-channel
Kondo problem by  Andrei and Wiegmann\cite{AFL}\cite{TW}.
There are many generalizations of the original Kondo model.
Among them one of the most important generalization is 
the multi-channel Kondo model\cite{AD}\cite{TW84}\cite{Fe}\cite{Af}.
In this model electrons have flavour (channel) degrees of 
freedom in addition to the spin (color) degrees of freedom
whereas impurity has only spin degrees of freedom.

On the other hand,
in the last few years, transport properties of 
interacting electrons  in one-dimensional (1D)
wire, which are also interacting with a single or double barrier, are studied 
by various methods\cite{FG}\cite{FLS1}.
These electrons are believed to be described by a Luttinger liquid, and
resemblance between 1D electron systems with double barrier, i.e., 
quantum dot, 
and the Kondo model was recognized in various aspects.  
For example, the number of electrons in the dot is severely restricted because of small
electric capacitance of the dot and only their spin degrees of freedom is a dynamical
variable.
Resonant tunneling of electrons through dot as a result of the Kondo type
interaction was predicted theoretically\cite{NgLee} and it was observed quite recently
by experiments\cite{exp}. 

There is another interesting electron system which is described by 1D Luttinger
liquid interacting with quantum barrier, i.e., quantum Hall states connected by
point contacts.
In most of 1D electron systems, 
localization effect is very strong
and backward scattering by impurities makes low-energy excitations are gapfull.
In this sense chiral Luttinger liquid\cite{W} 
is more stable than the right-left
symmetric one and experimentally easy to be observed.
Actually two quantum Hall edge states connected by a point contact
were investigated intensively, and theoretical studies in terms of 
chiral Luttinger liquid and experiments are in good agreement.
In this study exactly integrable model and its solutions play a very important
role\cite{FLS1}\cite{FLS2}.

In this paper, inspired by the above mentioned resonant tunneling 
through quatum dot in the (chiral) Luttinger liquid, 
we  shall introduce an impurity model which is a kind of generalization of
the multi-channel Kondo model.
In our model
both electrons and impurity have  $f$ species of flavour degrees of freedom 
in addition to the $N$ species of spin degrees of freedom.
%
Related model has been  already introduced for the other impurity problems
and studied using renormalization group and conformal field theory\cite{Y}.
In this paper integrabiblity of the model 
(with more general interactions) is proved and
we show that the model can be analysed using the Bethe ansatz solutions.

In the Kondo type models,  we are mostly interested in 
low-energy phenomena and excitations near the Fermi surface.
In most of studies, the linearized dispersion relation is employed
for electrons.
But in certain cases, to study the physical properties of the model correctly,
we must introduce second-derivative terms 
with momentum cutoff which is sent to infinity 
after the calculation\cite{AD}\cite{ZA} 
and additional electron-electron interaction terms in
the Hamiltonian.\footnote{For the model with finite cutoff, 
the model is integrable only when the electron-electron coupling
is some given value determined by electron-impurity coupling.
After removing the cutoff the model is integrable for 
arbitrary strength of the electron-electron coupling.}
Especially in the multi-channel Kondo model,
by introducing second-derivative terms and electron-electron interaction 
terms we get  
{\it bound states} of  electrons 
which are flavour singlet.
In this model flavour degrees of freedom of electrons influence the 
physical properties of impurity via these bound states.

Let us briefly review the integrability of the Kondo model with ultraviolet
cutoff following the studies in Refs.\cite{AD}\cite{ZA}.
In the ($SU(N)$) Kondo model, integrability is guarateed by the fact that
$S$ matrices are rational solutions to
the Yang-Baxter equations.
Electron-electron $S$ matrices of the model with 
cutoff $\Lambda$ are given by
\ba
&&
S_{ij}
=
\frac{\lam_i - \lam_j -ic P_{ij}}{
\lam_i - \lam_j -ic},
\ea
and
electron-impurity $S$ matrices are given as 
\ba
&&
S_{j0}
=
\frac{\lam_{j} -\alpha -ic P_{0j}}{\lam_{j} -\alpha -ic},
\label{mscat}
\ea
where 
$i,j = 1,\cdots, N_{e}$,
$S_{ij}$ ($S_{j0}$) acts to the tensor product of the spin spaces of the
 $N_{e}+1 $ electrons  and operates nontrivially only on p
spins of $i$-th and $j$-th electrons $ C^{N} \otimes C^{N}$
(resp.  of  electron $j$ and impurity), i.e.,
$P_{ij}$ exchange spin
coordinates of $i$-th and $j$-th electrons.
Prameter $\lam_{j}$ is given by $\lam_{j} = k_{j}/\Lam$,
where $k_{j}$ is the rapidities of electron $j$.  
The impurity  is denoted by the subscript $0$
and from (\ref{mscat}) $\alpha$ can be interpreted as the rapidity of  the 
impurity.
Hereafter we shall call the parameter $c$, in the above form of 
$S$ matrices, coupling constant of $S$ matrices.
Please notice that as the cutoff is sent to infinity
the above $S$ matrices become ordinary $S$ matrices of the
Kondo model\cite{AFL}\cite{TW}.
The above $S$ matrices  (1) and (2) satisfy the Yang-Baxter equations
\ba
&&
S_{ij} S_{i0} S_{j0} = S_{j0} S_{i0} S_{ij}, \nn \\
&&
S_{ij} S_{ik} S_{jk} = S_{jk} S_{ik} S_{ij}.
\ea

For the multi-channel Kondo model electron-electron $S$ matrices are 
given by
\ba
&&
S_{ij}
=
\frac{\lam_i - \lam_j -ic P_{ij}}{
\lam_i - \lam_j -ic}
\frac{\lam_i - \lam_j +ic \cP_{ij}}{
\lam_i - \lam_j +ic},
\label{mKS}
\ea
where these $S$ matrices act to the tensor product of the 
spin and flavour spaces of the $N_{e} + 1$ electrons,
$(C^{N} \otimes C^{f}) \otimes (C^{N} \otimes C^{f})$,
and $S_{ij}$ operates on the $i$-th and $j$-th  electrons, i.e.,
 $\cP_{ij}$exchanges  the flavour 
coordinate of the electrons $i$ and $j$.
Electron-impurity $S$ matrices are the same with (\ref{mscat}),
i.e., flavour parts of electron-impurity $S$ matrices 
are trivial.

This paper is organized as follows.
In Sect.2, we shall introduce a 1D electron-impurity model which is a kind of
generalization of the multi-channel Kondo model.
There are various parameters in the model.
It is shown that if these parameters satisfy certain relations, the model
is integrable and electron-electron and electron-impurity $S$ matrices
have factorized forms similar to  (\ref{mKS}). 
In Sect.3, Bethe ansatz equations are given.
In Sect.4, its relationship to quantum dot is explained, and 
section 5 is devoted for conclusion.

Thermodynamics of this model and detailed studies of
Bethe ansatz equation for arbitrary $f$ and $N$ 
will appear elsewhere\cite{T}.
Actually, one can consider new ``category" of integrable models whose
simplest example is the model given in this paper.
Some of generalization of the present model will be suggested 
in later discussions.
We shall report studies of them in future publications.
It is also interesting to study (solvable) field theory models
which correspond to the present model of the first quantization.
They are useful and can predict physical properties of 
resonant tunneling through
dot as varying gate voltage, etc.

\section{The integrability of the model}
\subsection{The model with finite cutoff}

In our model the electrons and impurity are
in the fundamental representation of spin $SU(N)$
and flavour $SU(f)$.
Impurity is located at the origin.
Wave functions of the model 
$f_{\{ \lam_j  \} , l}^{\{ \alpha_k \}, a}(\{x_i \})$
are functions of 
the coordinates of the conduction electrons $x_{1},\cdots , x_{N_{e}}$,
spin coordinates of the electrons (and impurity )
$\lam_{1}, \cdots ,\lam_{N_{e}}$ (resp. $l$)
and flavour coordinates of the electrons (and impurity)
$\alpha_{1}, \cdots ,\alpha_{N_{e}}$ (resp. $a$)
with 
$\{ \lam_{1}, \cdots ,\lam_{N_{e}}, l\} \in \{1,\cdots ,N\}$ and 
$\{ \alpha_{1}, \cdots ,\alpha_{N_{e}}, a\} \in \{1 , \cdots ,f\}$.

Model is defined by the following first 
quantized Hamiltonian with finite cutoff 
(hereafter we call this model A) 
\ba
&&
H 
= H_{ee} + H_{ei} + \tilde{H}_{ee} + \tilde{H}_{ei},
\ea
where
\ba
&&
H_{ee}
=
-i \sum_{j=1}^{N_{e}}  \partial_{j} 
 + \sum_{l<j}  \delta (x_l - x_j)(U_{c}P_{jl} - U_{f}\cP_{jl}) \nn \\
&&
H_{ei}
=
 \sum_{j=1}^{N_{e}} \delta (x_{j}) 
( J_{c} P_{0j} + J_{f} \cP_{0j}  ) \nn \\
&&
\tilde{H}_{ee}
=
- \frac{1}{2 \Lambda} \sum_{j=1}^{N_{e}} \partial^{2}_{j} \nn \\
\label{hamiltonian}
\ea
$H_{ee}$ consists of the kinetic term of the electrons with the linearized 
dispersion relation and electron-electron interactions.
$H_{ei}$ is the electron impurity interaction term.
$\tilde{H}_{ee}$ is the second-derivative term which introduces 
curvature in the dispersion relation
and  $\Lambda$ is the cutoff which will be sent to infinity
at the final stage of the calculation.

$U_c, U_f, J_{c}, J_{f}$ are all parameters of the present model,
and their physical meaning is explained in Sect.4 (see also Fig.2).
Note that for the infinite cutoff limit the model is integrable for
more general interactions (see Sect.2.2 ).


We shall diagonalize the Hamiltonian by the Bethe ansatz.
The Bethe ansatz wave functions for the model
with $N_{e}$ conduction eletrons are given by
\ba
&&
F_{\{ \lam_j  \} , l}^{\{ \alpha_k \}, a}(\{x_i \})
=
\cA \sum_{Q\in S_{N_{e}+1}} e^{\sum_{j=1}^{N_{e}} i k_{j} x_{j}} 
\xi_{\{ \lam_j  \} , l}^{\{ \alpha_k \}, a} (Q)
\theta(x_{Q}),
\label{bethe}
\ea
where $\cA$ is the antisymmetrizer,
 $Q$ runs over  all the permutations of electrons and impurity
and $\theta (x_{Q})$ is unity if $x_{Q(0)}<x_{Q(1)}  <\cdots <x_{Q(N_{e})}$
and is vanishing otherwise.

$\xi_{\{ \lam_j  \} , l}^{\{ \alpha_k \}, a} (Q)$ 
are the spin and flavour dependent amplitudes
in the region $Q$.
$\xi_{\{ \lam_j  \} , l}^{\{ \alpha_k \}, a} (Q)$ can be interpreted as 
the components of the vectors
$\xi (Q)$ in the spin and flavour space 
$\otimes_{j=1}^{N_{e}+1} (C^{N} \otimes C^{f})$.
$\xi (Q)$'s in different regions are related by the $S$ matrices as 
\ba 
&&
\xi ((i,j)Q)
=
S_{i j} \xi (Q),
\ea
where $ Q = (Q(0), \cdots , Q(l)=i,Q(l+1)=j, \cdots , Q(N_{e}))$
and  $ (i,j)Q = (Q(0), \cdots , Q(l+1),Q(l), \cdots , Q(N_{e}))$.
Consistency of the Bethe ansatz wave functions is 
guaranteed if the $S$ matrices satisfy the Yang-Baxter equations.


The electron-electron $S$ matrices are factorized to the 
spin part and flavour part,
\ba
&&
S_{ij} 
= S_{ij}^{c} S_{ij}^{f},
\ea
where 
\ba
&&
S_{ij}^{c}
=
\frac{\lam_i - \lam_j -i(U_{c}+U_{f}) P_{ij}}{
\lam_i - \lam_j -i (U_{c}+U_{f})} \nn \\
&&
S_{ij}^{f}
=
\frac{\lam_i - \lam_j + i (U_{c}+U_{f}) \cP_{ij}}{
\lam_i - \lam_j +i (U_{c}+U_{f})},
\label{newS}
\ea
and $\lam_{j} = k_{j}/\Lam$.
Note that for the ordinary Kondo Hamiltonian, where only first-derivative
terms in $H_{ee}$ exist without the 
second-derivative terms in $\tilde{H}_{ee}$ and the electron-electron 
interaction, 
electron-electron $S$ matrices cannot be
determined by the Hamiltonian itself.
$S$ matrices are determined by demanding that $S$ matrices
satisfy the Yang-Baxter equations.
But in the present case, $S$ matrices are determined by the
Hamiltonian uniquely, and by removing the cutoff we reconstruct 
the well known electron-electron $S$ matrices of the 
Kondo model\cite{AD}\cite{ZA}.

Note also that with any reasonable choice of interactions
the spin part and flavour part of the $S$ matrices
have the same form up to the sign of the 
coupling constant term $(U_{c}+U_{f})$ in (\ref{newS}).

Let us turn to the electron-impurity S matrices.
For the electron-impurity interaction of the 
form given by Hamiltonian (\ref{hamiltonian})
electron-impurity $S$ matrices are factorized.
Furthermore,  if the coupling constants satisfy the equation
\ba
&&
J_{f} = - J_{c},
\label{condition2}
\ea
then the coupling constants of the $S$ matrices in the spin part
and in the flavour part are same
(this condition is necessary for the $S$ matrices to satisfy
the Yang-Baxter equation because
the coupling constants of the $S$ matrices in the spin part 
and in the flavour part of the electron-electron $S$ matrices
are the same).


When the coupling constants satisfy  
Eq.(\ref{condition2}),
electron-impurity $S$ matrices are given by
\ba
&&
S_{j0} 
=
S_{j0}^{c} S_{j0}^{f},  \\
&&
S_{j0}^{c}
=
\frac{\lam_{j} -\alpha -ic P_{0j}}{\lam_{j} -\alpha -ic}, \nn \\
&&
S_{j0}^{f}
=
\frac{\lam_{j} -\alpha + ic \cP_{0j}}{\lam_{j} - \alpha + ic}.
\label{electronimpurity}
\ea
Here $\alpha$ and $c$ are  functions
of coupling constants $(J_{c},J_{f})$,
and for the model to be integrable
electron-electron coupling constants $(U_{c},U_{f})$
must satisfy the condition 
\ba
&&
U_{c}+U_{f} = c.
\ea

Notice  that there is another case in which the 
$S$ matrices of the model is
the solution to the  Yang-Baxter equations,
that is, the coupling constants satisfy 
the following conditions
\ba
&&
 J_{f} =0 ~~~\mbox{and} ~~J_{c}~ \mbox{is finite},
\label{multichannel}
\ea
(or the case with the spin and flavour exchanged in the above).
In the above case, the model is nothing but the multi-channel 
Kondo model.

Integrable region of the model 
in the space of electron-impurity coupling constants 
is given by two regions (crossing at the trivial point 
$J_{c}=J_{f}=0$) characterized by 
Eqs.(\ref{condition2}) and (\ref{multichannel}).

%
\subsection{The model with infinite cutoff}

We shall study another model (which we call model B) whose
Hamiltonian is given by
\ba
&&
H
= 
-i \sum_{j=1}^{N_{e}}  \partial_{j} 
 + \sum_{l<j}  \delta (x_l - x_j)(U_{c}P_{jl} - U_{f}\cP_{jl}) \nn \\
&&
+ \sum_{j=1}^{N_{e}} \delta (x_{j}) 
( J_{cf} P_{0j} \cP_{0j} + J_{c} P_{0j} + J_{f} \cP_{0j} +J )
\label{ihamiltonian}
\ea
This model is obtained by taking the cutoff to infinity
in the model A and also the electron-impurity interactions
of the model B are more general.

To obtain the correct results,
we must first study the model A and classify the complex
solutions (string) of the Bethe ansatz equations, and
then study the model B for more general cases.

In this model electron-electron $S$ matrices are given by
\ba
&&
S_{ij} 
=
P_{ij} \cP_{ij},
\label{SmodelB}
\ea
and electron-impurity $S$ matrices have the following form 
\ba
&&
S_{j0}
=
f P_{j0}\cP_{j0} + g P_{j0} + h \cP_{j0} + l,
\ea
where $f,g,h$ and $l$ are complicated functions of 
$J_{cf},J_{c},J_{f},J$, and 
for arbitrary  coupling constants
the model is integrable (at least) formally (i.e., $S$ matices satisfy the 
Yang-Baxter equations).
But in practice only when the electron-impurity $S$ matrices
are factorized to the spin part and  flavour part,
we can diagonalize the eigenvalue problem which appears by  
imposing the periodic boundary condition on the wave function.
The condition that $S$ matrices are factorized is given by\footnote{
In this paper factorization means that $S$ matrices are 
factorized to spin part and flavour part.}.
\ba
&&
J_{cf}^{3} - J_{cf} (J_{c}^{2} + J_{f}^{2}) 
+ J_{cf} ( - 4  - J^{2})
+ 2 J J_{c} J_{f} = 0.
\label{bcondition}
\ea

When the coupling constants satisfy Eq.(\ref{bcondition})
electron-impurity $S$ matrices are given by
\ba
&&
S_{j0} 
=
S_{j0}^{c} S_{j0}^{f},
\ea
\ba
&&
S_{j0}^{c}
=
\frac{-\alpha_{c} -ic_{c} P_{0j}}{ -\alpha_{c}-ic_{c}} \nn \\
&&
S_{j0}^{f}
= 
\frac{ -\alpha_{f} + ic_{f} \cP_{0j}}{- \alpha_{f} + ic_{f}}.
\ea

In the model B even if $c_c \neq c_f$,
the Hamiltonian can be diagonalized using Bethe ansatz
as seen  from the form of the electron-electron
$S$ matrices (\ref{SmodelB}).
This broadness of the interaction is important 
when the model is applied to experimentally accessible
physical systems like the resonant tunneling 
in the quantum dot (see discussion in Sect.4).

There is a simple nontrivial solution
to the factorizability condition of the electron-impurity
$S$ matrices, i.e., Eq.(\ref{bcondition}).
That is 
\ba
&&
J_{cf} = J =0.
\label{ssolution}
\ea
The electron-impurity interactions of the model A
have the form of this solution
except that for the model A there exist
additional condition for the model to be
integrable that is $J_c = - J_f$.
Parameter region in which the model A is integrable
and the special solution for the model B is given  in  
Fig. 1.

For the solution (\ref{ssolution}) $S$ matrices  are expressed
explicitly as
\ba
&&
S_{j0}^{c}
=
\frac{2 + i \frac{8 J_{c}}{J_{c}^{2} -J_{f}^{2} -4}P_{0j}}
{2 + i \frac{8 J_{c}}{J_{c}^{2} -J_{f}^{2} -4}} \nn \\
&&
S_{j0}^{f}
=
\frac{2 - i \frac{8 J_{f}}{J_{c}^{2}- J_{f}^{2} + 4}\cP_{0j}}
{2 + i \frac{8 J_{f}}{J_{c}^{2}- J_{f}^{2} + 4}}.
\ea

\section{Bethe ansatz equations of the models}

In this section we shall diagonalize the Hamiltonian
of the model for the special case $N=f=2$ by the Bethe ansatz.

Imposing the periodic boundary conditions for the
wave function (\ref{bethe}),
we obtain  eigenvalue problems
\ba
&&
e^{ik_{j} L} \xi (Q)
=
Z_{j} \xi  (Q),
\ea
where
\ba
&&
Z_{j}
=
S_{j j-1} \cdots S_{j 1} S_{j 0} S_{j N} \cdots S_{j j+1}
\ea
are matrices which operate 
to vectors in $\otimes_{j=1}^{N_{e}+1} (C^{N} \otimes C^{f})$
and each factor comes from the scattering of $j$-th electron by another electron
and the impurity.

Since $P_{ij}$ and $\cP_{ij}$ commute with each other,
$Z_{j}$ can be factorized to spin part and flavour part as
$Z_{j} = Z_{j}^{f} Z_{j}^{c}$,
\ba
&&
Z_{j}^{f} 
=
S_{j j-1}^{f} \cdots S_{j 1}^{f} S_{j 0}^{f} 
S_{j N}^{f} \cdots S_{j j+1}^{f} \nn \\
&&
Z_{j}^{c}
=
S_{j j-1}^{c} \cdots S_{j 1}^{c} S_{j 0}^{c} 
S_{j N}^{c} \cdots S_{j j+1}^{c}.
\ea 
Eigenvectors of $Z_{j}$ are simultaneous 
eigenvectors of $Z_{j}^{c}$ and $Z_{j}^{f}$, and
then we can diagonalize the spin and flavour degrees of freedom
independently\cite{TW}\cite{A}.
Operators $Z_{j}^{c}$ and $Z_{j}^{f}$ are 
transfer matrices of the inhomogeneous XXX model
at the special values of the rapidities
and they can be diagonalized by the algebraic Bethe ansatz\cite{TW}\cite{A}.

Resultant Bethe ansatz equations in the model A are given by
\ba
&&
e^{ik_{j}L} 
=
\prod_{\gamma = 1}^{M} \frac{\omega^{c}_{\gamma} - \lam_{j} - \frac{ic}{2}}
                       {\omega^{c}_{\gamma} - \lam_{j} + \frac{ic}{2}}
\prod_{\gamma = 1}^{\bar{M}} 
     \frac{\omega^{f}_{\gamma} - \lam_{j} + \frac{ic}{2}}
                       {\omega^{f}_{\gamma} - \lam_{j} - \frac{ic}{2}}, \nn \\
&&
\prod_{\delta (\neq \gamma )} 
               \frac{\omega^{c}_{\delta} - \omega^{c}_{\gamma} -i c}
             {\omega^{c}_{\delta} - \omega^{c}_{\gamma} + i c}
= 
\prod_{i=1}^{N}
\frac{\omega^{c}_{\gamma} - \lam_{i} + \frac{ic}{2}}
  {\omega^{c}_{\gamma} - \lam_{i} - \frac{ic}{2}}
\frac{\omega^{c}_{\gamma} - \alpha +\frac{ic}{2}}
   {\omega^{c}_{\gamma} - \alpha - \frac{ic}{2}}, \nn \\
&&
\prod_{\delta (\neq \gamma )} 
               \frac{\omega^{f}_{\delta} - \omega^{f}_{\gamma} -i c}
             {\omega^{f}_{\delta} - \omega^{f}_{\gamma} + i c}
=
\prod_{i=1}^{N}
\frac{\omega^{f}_{\gamma} - \lam_{i} + \frac{ic}{2}}
  {\omega^{f}_{\gamma} - \lam_{i} - \frac{ic}{2}}
\frac{\omega^{f}_{\gamma} - \alpha +\frac{ic}{2}}
   {\omega^{f}_{\gamma} - \alpha - \frac{ic}{2}},
\label{baequation}
\ea
where $\omega^{c}_{\gamma}$'s are spin rapidities,
  $\omega^{f}_{\gamma}$'s are flavour rapidities
and $M$($\bar{M}$) is the number of down(up) spins of the flavour $SU(2)$.
Energy of the state is give by
\ba
&&
E
=
\sum_{j=1}^{N_{e}}\Big( k_{j} + \frac{1}{2 \Lam} k_{j}^{2}\Big).
\ea

The Bethe ansatz equations (\ref{baequation}) are 
related to those for the ordinary Kondo model.
The spin part and flavour part of the equations (\ref{baequation}) have the 
same form with the spin part of the ordinary Kondo model (with the cutoff).
Then the classification of the string solution for the
Bethe ansatz equations is very similar to that of the Kondo model.
Classification of string solutions and study of the finite temperature
thermodynamics will appear elsewhere\cite{T}.


\section{Quantum dot}

In this section, we explain the relationship between the present model (with $N=f=2$)
and electron transport phenomena through quantum dot.
There are two cases to which our model is relevant,
one is the system of quantum dot in  
1D interacting electrons (Luttinger liquid)
and the other is the double-layer quantum Hall edges which are 
contacted by a quantum dot.

\subsection{Electrons and dot in one-dimensional wire}

For the system of quantum dot in the Luttinger liquid,
we shall interpret the flavour degrees of freedom in the present model as
chirality of  electrons.
Strictly speaking, signature of the linear momentum term in the 
electron Hamiltonian depends on the chirality of the electron, i.e.,
\ba
&&
-i \sum_{j=1}^{N_{e}} \alpha_{j} \partial_{j},
\ea
where $\alpha_{j} = +1$($-1$) 
if $j$-th electron is rightmover(leftmover).
However we think that chirality
does {\em not} give substantial influences in the present case because 
interaction between electrons and the dot  occures 
only at the point $x=0$\footnote{Chirality, of curse, plays an
important role in the localization phenomenon by random impurities.
Becase of this, chiral Luttinger liquid can be realized 
in real materials and experiments
more easily.}(see similar discussion in Refs.\cite{FLS1}\cite{FLS2}).
Explicitly in qunatum field language left and right-moving fermions,
$\psi_L(x)$ and $\psi_R(x)$, can be regarded as flavour-two-component
left mover by $\psi_1(x)=\psi_L(x), \; \psi_2(x)=\psi_R(-x)$.
Flavour change at $x=0$ is therefore nothing but reflection or
backscattering at $x=0$ (see Fig.2).

As we explained in the introduction, 
charge of quantum dot is strictly restricted
by small electric capacitance of the dot.
Therefore if electron number in the dot is restricted to be odd, 
only spin degrees of freedom
of the dot is dynamical and the dot can be regarded a 
spin-${1 \over 2}$ impurity as in the Kondo model.

Let us study the above system in more detail and ``derive" 
the Hamiltonian in this paper from microscopic point of view.
Coulomb interaction in the dot is described by the term like
\begin{equation}
H_C={U_{\rho} \over 2}\Big(n_{\rho}-n_{0\rho}\Big)^2,
\label{Coulomb}
\end{equation}
where $U_{\rho} \sim e^2/l_d$ with linear magnitude of the dot $\l_d$,
$n_{\rho}$ is the electron number in the dot and  $n_{0\rho}$ is its 
mean value controlled by gate voltage. 
Let us consider the case in which $n_{0\rho}$ is an odd integer and 
lower energy levels are filled with spin up and down electrons, and therefore
we focus on the the lowest unfilled energy level (LUEL) inside the dot.

It is helpful to consider descrete version of the above system 
and let us denote electron operator in the LUEL in the dot  by $c_{0\sigma}$
where $\sigma=\pm 1$ is the spin index.
Then the Hamiltonian is given as
\begin{equation}
H_{DV}=H_0+H_{t'}+H_C,
\label{discrete}
\end{equation}
\begin{eqnarray}
H_0 &=& -t \sum_{\sigma}\Big[\sum_{i \geq 1}(c^{\dagger}_{i\sigma}c_{i+1\sigma}
+\mbox{H.c.}) + \sum_{i \leq -2}(c^{\dagger}_{i\sigma}c_{i+1\sigma}
+\mbox{H.c.})\Big] , \nonumber    \\
H_{t'}&=& -t'\sum_{\sigma}(c^{\dagger}_{0\sigma}c_{1\sigma}+c^{\dagger}_{-1\sigma}c_{0\sigma}
+\mbox{H.c.}),  \label{H0}
\end{eqnarray}
where $i$ is the site index and $H_C$ is given by (\ref{Coulomb}) with
$n_{\rho}=\sum_{\sigma}c^{\dagger}_{0\sigma}c_{0\sigma}$
and $n_{0\rho}=1$.
In the present case $U_{\rho} \gg t'$ and the electron transmission between the dot
and the leads is treated by perturbation.
In the dot, the double-occupied state has large energy $U_{\rho}$
compared to the single-occupied states, and therefore we shall focus on the
subspace of single occupancy in the LUEL.
Then it is rather straightforward to derive an effective Hamiltonian for
the subspace by mean of the perturbative calculation for $H_{t'}$, e.g., 
\begin{eqnarray}
 _R\langle \downarrow| _0\langle \uparrow |H_{t'}(E_0-H_0-H_C)^{-1}H_{t'}
    |\uparrow\rangle_0 |\downarrow\rangle_R&=& -2 t'^2/U_{\rho}, \nonumber  \\
 _R\langle \downarrow| _0\langle \uparrow |H_{t'}(E_0-H_0-H_C)^{-1}H_{t'}
    |\downarrow\rangle_0 |\uparrow\rangle_R&=& 2 t'^2/U_{\rho},  \nonumber  
\end{eqnarray}
etc., where intermediate state of the dot is the double-occupied state 
$c^{\dagger}_{0\uparrow}c^{\dagger}_{0\downarrow}|0\rangle_0$
and $|\cdot\rangle_R$ is a state of free electron in the right side of the dot.
Similarly conduction electron can be transferred across the dot with or
without flipping its own spin and that of the dot.
Therefore there appear Kondo-type interactions between spin of the dot and
that of conduction electrons.    
In the first qunatization, these interactions between spin degrees of freedom of the dot
and conduction electrons are described by the terms proportional to the
spin and flavour exchange operators $P_{0j}$, ${\cal P}_{0j}$ etc.
    
The various parameters in the Hamiltonian of the present model 
(\ref{ihamiltonian})
have the following physical meanings in the electron transport properties
through quantum dot (see Fig.2).
$U_c$ is the spin flip interaction between electrons, whereas $U_f$ is 
the strength of {\it backscattering} of electrons. 
For clean samples, these parameters become small.
Similarly $J_c$ represents the electron transmission through the dot
with spin flip.
Flavour exchange operator ${\cal P}_{0j}$ is related to flavour angular momentum 
operators of the dot $\vec{S}_f$ and electron flavour current
$\psi^{\dagger}\vec{\tau}\psi$ as
\begin{equation}
{\cal P}_{0j} \propto \vec{S}_f\cdot \psi^{\dagger}
\vec{\tau} \psi(0)+C,
\nonumber  
\end{equation}
where $C$ is some constant.
As explained above, flavour change at $x=0$  $\psi_1 \leftrightarrow \psi_2$ is
$\psi_L(x) \leftrightarrow \psi_R(-x)$.
Therefore $J_f$ and $J_{cf}$ correspond to electron ``backscattering"  
by the dot with and without spin flip, respectively.
In experiments these backscatterings by dot are controlled
by varying, e.g., gate voltages, in order to observe resonant tunnelling.  
Actually condition of resonant tunneling is $J_f=0$\cite{KF}, i.e.,
vanishing of electron backscattering by potential barrier by the dot.

We should also remark here that in order to describe 1D electron system with
quantum dot faithfully flavour of the dot must be sufficiently large.
Otherwise backscatterings of electrons by the dot alternate in
direction.
We can employ cyclic representation of qunatum group $SU(2)_q$
for flavour degrees of freedom of impurity for the above purpose.
More systematic studies are possible by using (solvable) field theory
model corresponding to the present model in the first quantization.
This problem is under study and results will be reported
in future publication\cite{IT}.

\subsection{Quantum Hall edges in double-layer system} 

Another related system to application of the present model is quantum Hall state(QHS) in
a double layer electron system and spin-singlet QHS.
In these cases, low-energy excitations are chiral edge states with (pseudo-)spin
degrees of freedom.
If two edges states in double-layer QHS are connected by dot, which is also 
double-layer system, two edge states are labeled by flavour indices and
layers are labeled by pseudo-spin indices
(we assume here that real spins of electrons are polarized
by an external magnetic field).
It is well known that the pseudo-spin description is useful for study of low-energy
excitations in the double-layer QHS\cite{DLQHS}.
In double-layer dot, state with a ``movable" electron in the upper(lower) layer
has pseudo-spin $+{1 \over 2}(-{1 \over 2})$ (see Fig.3).
As in the previous case, Coulomb interaction energetically prefers 
the single-occupied states in the quantum dot, and because of that
Kondo-type interactions effectively appear in the low-energy subspace.
All relevent interactions between fermions and dot in the above
system are obviously included in the Hamiltonian.
Here again field theory model is possible to be constructed for studying this system
and especially its bosonized form is quite usefull for
the calculation of conductivity, etc. , as in the $\nu={1 \over 3}$ single
layer case\cite{FLS2}.
There are more parameters in the double layer case compared with
the single-layer case and therefore 
we can expect interesting phenomena.
This problem is under study and we shall report results in future publication.

\section{Conclusion}

In this paper we have introduced the Kondo type integrable
impurity model with spin and flavour degrees of freedom
in both electrons and impurity,
and we have diagonalized the Hamiltonian by the Bethe ansatz for the case of 
$N=f=2$.
Among the low-energy excitations,
there exists what we call {\it flavouron}
in addition to the holon and spinon excitations.
Detailed study of the model for arbitrary $N$ and $f$ and 
finite-temperature thermodynamics of the model will appear 
elsewhere\cite{T}.

If we interpret impurity as quantum dot,
flavour degrees of freedom in 
impurity should not be the spin $1/2$ representation of $SU(2)$, as we
explained in Sect.4.
In a faithful model which describes quantum dot, flavour degrees of freedom
of the impurity is given by a cyclic representation of $ SU(2)_{q}$ 
or spin $j/2$ representation of $SU(2)$ with $j \rightarrow \infty$.
Then it is very interesting to study the model in which
spin degrees of freedom of both electrons and the impurity 
belong to the fundamental representation of $SU(N)$
whereas flavour degrees of freedom of the impurity
belong to the spin $j/2$ representations of 
$SU(2)$\cite{T}\cite{AFL}\cite{TW}.

The model in which flavour degrees of freedom is 
in the cyclic representation of $SU_{q}(2)$  for the impurity 
can be more easily  studied by bosonizing the
model and map the problem to the integrable field theory with 
boundary\cite{GZ}\cite{IT}\cite{Y}.
To study the transport properties of the model, the boson model
is more powerful than the original model\cite{FLS1}.

There are some generalizations of the model in this paper.
In this paper only isotropic case (corrsponding to the
rational solution of Yang-Baxter equations)
was studied.
For the application to the resonant tunneling in quantum dot,
study of anisotropic case\cite{TW} is interesting,
and study of the model in terms of the boundary conformal field theory
is also important\cite{Af}.
As we stated in Sect.4, 
in order to interpret flavour in our model as chirality 
the kinetic term which is linear in momentum should be modified.
This problem is under study.

The model is related to the resonant tunneling through  
quantum dot in (chiral) Luttinger liquid.
There is another interesting system of
quantum dot which is described by an  integrable impurity model with 
flavour degrees of freedom, that is, {\em multi-channel} quantum dot.
We found that the following Hamiltonian which represents  
this phenomenon is integrable,
\ba
&&
H
=
-  \sum_{j=1}^{N_{e}} \partial^{2}_{j} 
 + \sum_{l<j}  \delta (x_l - x_j)(U_{c}P_{jl} - U_{f}\cP_{jl})  \nn \\
&&
+
 \sum_{j=1}^{N_{e}} \delta (x_{j}) 
\{ J_{c} P_{0j} + J_{f} \cP_{0j}  \},
\ea
where there are some costraints between the coupling constants
(there is relation between electron-electron coupling constant 
and electron-impurity coupling constant).
Detailed will be reported in near future.

\vspace{12pt}
 
\noindent
\begin{center}
{\bf Acknowledgement}
\end{center}

One of the authors (O.T.) would like to thank to
the members for the seminar of Kuniba laboratory 
especially Dr. J. Suzuki for the discussions.

%
%
%

\newpage 

\begin{figure}[h]

\begin{scriptsize}
\begin{picture}(330,200)

\thicklines
\put(0,75){\line(1,0){300}}
\thinlines
\multiput(125,0)(0,5){35}{\line(0,3){1}}
\thicklines
\put(50,0){\line(1,1){180}}

\put(20,20){model A}
\put(260,76){$J_{c}$}
\put(130,160){$J_{f}$}
\end{picture}

\end{scriptsize}
\caption{The region of the model to be integrable (if $J_{cf}=J=0$).
For the model A only on the solid lines $J_{c}=-J_{f}$ and 
$J_{f}=0$ (multi-channel Kondo model) the model is integrable.
For the model B on the entire space the model is integrable.}
\end{figure}
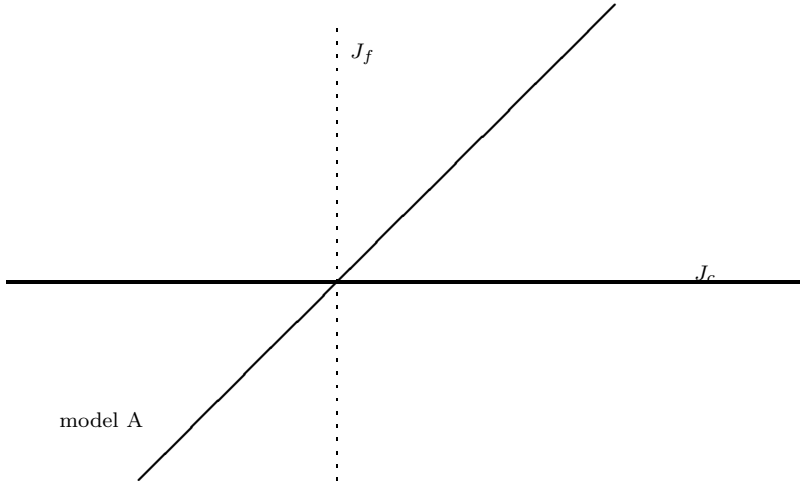             
\begin{figure}
\caption{The relation between Hamiltonian (16)
for $f=2$
and the resonant tunneling of Luttinger liquid in quantum dot.
The flavour degrees of freedoms are interpreted as the right or left movers.}
\end{figure}

\begin{figure}
\caption{Two layer quantum Hall states are connected by 
the quantum dot which has also two layers.
}
\end{figure}

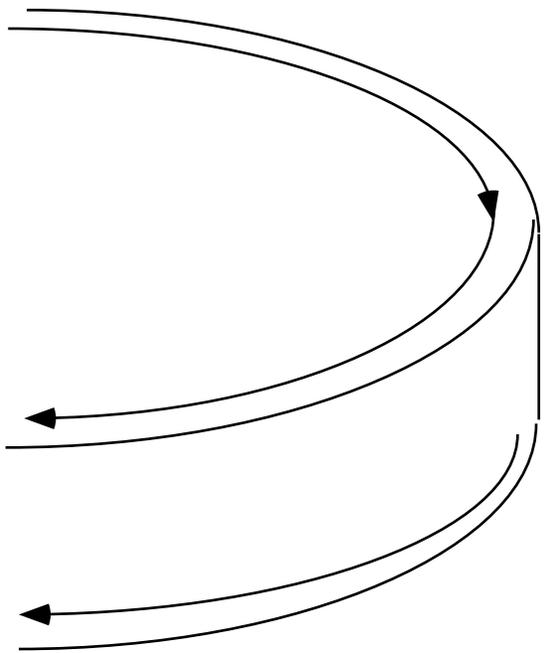
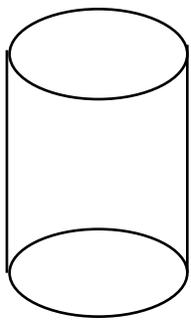
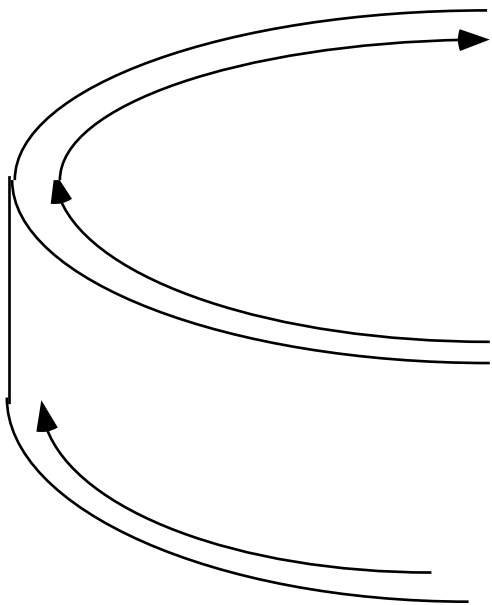
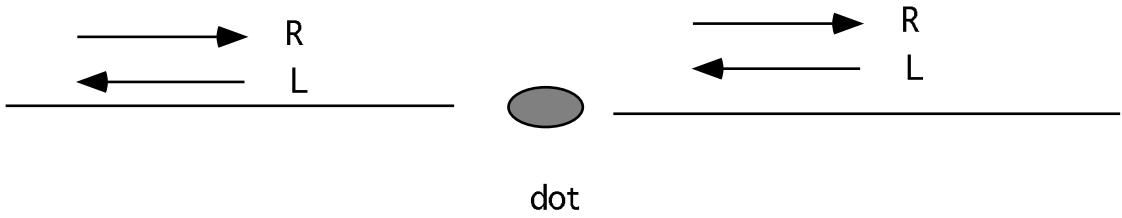
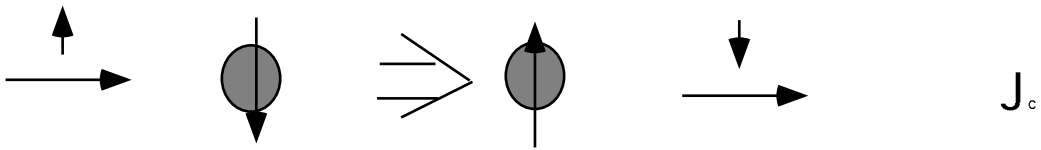
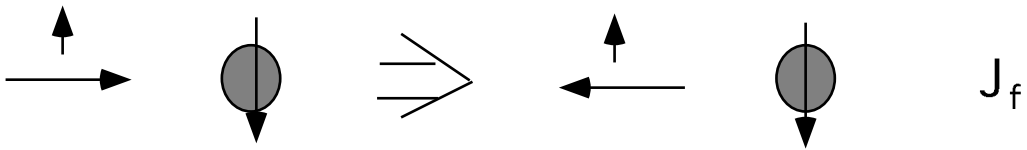
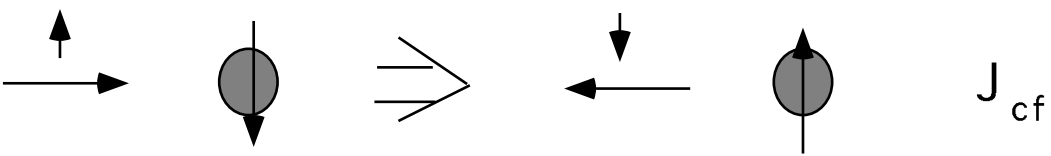
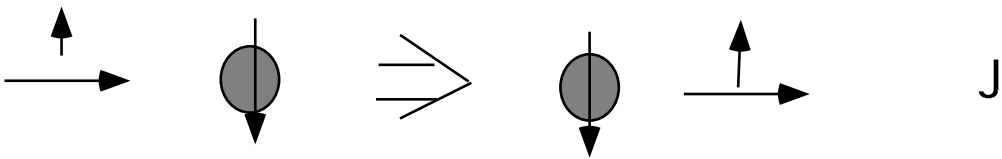
\end{document}